\documentclass[pra,reprint,superscriptaddress,showpacs]{revtex4-2}

\usepackage[T1]{fontenc}
\usepackage{booktabs}
\usepackage{mathptmx}
\usepackage{graphics}
\usepackage{multirow}
\usepackage{subfigure}
\usepackage{graphicx}
\usepackage{amsmath}
\usepackage{makecell}
\usepackage[colorlinks,
linkcolor=blue,
anchorcolor=red,
urlcolor=blue,
citecolor=blue]{hyperref}
\usepackage{threeparttable}
\begin{document}
	
	\title{\textbf{Re-understanding of the deformation potential constant in the single crystal silicon }}%
	
	\author{Aijun Hong}%
	\email[Corresponding author: ]{haj@jxnu.edu.cn, 6312886haj@163.com}
    	
    \affiliation{School of Physics, Communication and Electronics, Jiangxi Normal University, Nanchang 330022, China}

   \author{Feng Sun}%
   \affiliation{School of Education, Nanchang Institute of Science and Technology, Nanchang 330108, China}

	\date{\today}%
\begin{abstract}
The mobility formula based on deformation potential (DP) theory is of great importance in semiconductor physics. However, the related calculations for the DP constant are controversial. It is necessary to redo in-depth and comprehensive research on the mobility of single crystal silicon and the related parameters such as the effective mass and the DP constant. In this work the mobility effective mass is defined and a method based on the first principles is presented to evaluate the correction of the DP constant. It is found that the effective mass is closer to experimental data and the correction of the DP is a negligible value of about 0.3 eV being far lower than $\sim$10 eV calculated by other methods. Using these parameters, we obtain the mobilities of the single crystal silicon in reasonable agreement with the experimental values.

\end{abstract}
\maketitle

\section{Introduction}
Since the deformation potential (DP) theory was proposed by Bardeen and Shockley in 1950 \cite{rr1}, it has been applied to describe the interaction between electron and phonon \cite{rr45,rr46,rr47} and evaluate transport properties of semiconductors successfully \cite{rr2,rr43,rr44}. In particular, its combination with density functional theory (DFT) provides a favorable mean for explaining experiments and predicting new materials in the thermoelectric community. However, there are still some disagreements about the calculation for the DP constant of bulk materials. The focus of disagreements is that no precise reference energy level is found for obtaining the absolute band energy of the different deformation structures.

As is well known, the effective one-electron Schr\"{o}dinger (or Dirac) equation with respect to the eigenstates and eigenvalues ($\Psi_{i}$ and $\varepsilon_{i}$) is expressed as \cite{rr17}
\begin{equation}\label{e1}
  [K_{op}+V_{eff}(\textbf{r})]\psi_{i}(\textbf{r})=\epsilon_{i}\psi_{i}(\textbf{r}),
\end{equation}
where the effective Coulomb potential operator is the sum of the exchange correction potential ($\mu_{xc}$) and Coulomb potential ($V_{c}$)
\begin{equation}\label{e2}
  V_{eff}(\textbf{r})=\mu_{xc}(\textbf{r})+V_{c}(\textbf{r}).
\end{equation}
The $V_{c}$ is given by
\begin{equation}\label{e3}
  V_{c}(\textbf{r})=\int\frac{\rho(\textbf{r}')}{|\textbf{r}-\textbf{r}'|}d\textbf{r}'-\sum_{\alpha}\frac{Z_{\alpha}}{|\textbf{r}-\textbf{R}_{\alpha}|}.
\end{equation}
Obviously, there are the Coulomb singularities at the electron and nucleus sites ($\textbf{r}'$ and $\textbf{R}_{\alpha}$). In order to solve the Eq.~(\ref{e1}), there are many methods proposed to deal with the singularities. However, it is impossible to avoid introducing an uncertain constant when solving the Poisson's equation in any case. In fact, introducing the uncertain constant ignores the effects of both volume deformation and atomic species on the periodic potential. As a result, it makes the band energies of different materials and of the same material with different volumes, unable to be compared quantitatively, although the influence on the total energy can be canceled by special mathematical processing \cite{rr19,rr20,rr21}.

For the calculation of the DP constant of one material, it is not necessary to consider the influence of atomic species. Therefore, finding a physical quantity that is not affected by the volume deformation as the reference energy level is the key to calculating the DP constant. At present, there are average potential \cite{rr22,rr23,rr24,rr25} and core state level \cite{rr16,rr11} alignments to define the reference energy level. The former is also known as the supercell method, where the band structures are aligned according to the average potential from the core electrons in the supercell composed of the compressed and expanded parts. The latter is based on the assumption that the energy level of the core electron is hardly affected by the volume deformation.

Usually, the core state level alignment method is adopted to explore thermoelectric properties due to the requirement of less computational resources. In fact, the core state levels are sensitive to the volume deformation, and thus have non-negligible DP values \cite{rr16}. Consequently, the DP constants corrected by the core state level alignment are small and thus lead to much high mobility deviating from the experimental results. The disagreement between theory and experiment cannot be completely attributed to the theoretical calculation, because the recognized accurate experimental method for determining the DP constant has not been proposed.

Experimentally, the present determinations of the DP constant are only performed indirectly with a significant amount of assumptions. Therefore, the experimental DP constant for the same material has great differences. Taking GaAs as an example, Nolte et al. used the impurity level as the reference level to obtain the DP value of -9.3 eV for the conduction band minimum (CBM) \cite{rr3}, however, its value is 7.0 eV from the mobility measurements at low temperature \cite{rr4}. Interestingly, a value of $16.0 \pm0.5$ eV for CBM is obtained from high-temperature mobility \cite{rr7}. Moreover, Pfeffer et al. \cite{rr5} inferred from the optical absorption in Se-doped GaAs bulk \cite{rr6} that the DP value is as high as 15.7 eV. A much higher value of 20 eV is accepted and adopted in the Supporting Information of the recent literature \cite{rr12}. In a word, the experiment has not so far given satisfactory results about the DP constant for GaAs.

Theoretically, the DP constant of a material also varies greatly. In the original literature of DP theory \cite{rr1}, Bardeen and Shockley first deduced 11.3 eV for the DP value of the valence band maximum (VBM) of cubic silicon from measured mobility data in the polycrystalline silicon. Latter, the values -10.2 eV \cite{rr8}, -7.9 eV \cite{rr9}, and 2.05 eV \cite{rr11} are proposed by different researchers. It was claimed in some literatures \cite{rr24,rr16,rr14,rr15} that the obtained $\sim$1.5 eV along the [100] direction is very consistent with the experimental value $1.8\pm0.7$ eV \cite{rr13} for the p-type silicon. However, it is puzzled that we have not found relevant experimental data and only see the value of DP constants ($3.3 \pm0.7$ eV) for the n-type silicon in Ref. \cite{rr13}. Furthermore, it is worth mentioning that the experimental data is obtained in the heavily doped n-type silicon with the high concentration of $5\times10^{21}cm^{-3}$ and its Fermi level is located above the CBM. Therefore, the measured DP constant is for the Fermi level above the CBM, however, the theoretical DP constant is for the CBM \cite{rr24}. This implies that the theory and experiment are not consistent, even may deviate.

In the present work we first redefined the relevant physical parameters in the mobility formula, based on the analysis and summary of previous theoretical and experimental data on the DP constant. Then we present a strategy to evaluate the correction of the DP constants of materials with the aid of all-electron first-principles approach. Finally, we employ the modified mobility formula to calculate the mobility of the single crystal silicon for verifying the effectiveness of the strategy.

\section{Methodology}
According to the DP theory, the carrier mobility $\mu$ of non-polar crystals with the properties of isotropy and non-degeneracy is expressed as \cite{rr1}
\begin{equation}\label{ee4}
  \mu_{e(h)}=\frac{2\sqrt{2\pi}e\hbar^{4}\rho v^2}{3(k_{B}T)^{3/2}m^{5/2}_{e(h)}(\lambda_{e(h)})^2},
\end{equation}
where $k_{B}$, $e$ and $\hbar$ represent, respectively, the Boltzmann constant, the electronic charge and the reduced Planck constant, and $\rho$,$v$, $m_{e(h)}$ and $\lambda_{e(h)}$ are the mass density, the longitude acoustic velocity, the effective mass and DP constant for the n- or p-type (electron/hole) carrier (namely, for the CBM or valence band maximum (VBM)).

The $v$ of the isotropic polycrystal is govern by \cite{rr27}
\begin{equation} \label{ee5}
{{\textit{v}}} = {{\left( \frac{3{{B}}+4G}{3\rho } \right)}^{1/2}},
\end{equation}
where $B$ and $G$ are the bulk modulus and the shear modulus, respectively. Sometimes the following expression, neglecting the shear strain contribution to the longitude acoustic velocity, is adopted.
\begin{equation} \label{ee6}
{{\textit{v}}} = {{\left( \frac{{{B}}}{\rho } \right)}^{1/2}}.
\end{equation}
This is reasonable because small shear strain does not induce the volume change and therefore cannot generate effective DP value.

The longitude acoustic velocity $v^\beta$ along the $\beta$ direction in the single crystal is approximatively expressed by the elastic constant $c_\beta$ and the mass density,
 \begin{equation} \label{ee7}
v^\beta=\sqrt{\frac{c^\beta}{\rho}}.
\end{equation}

For the anisotropic single crystal, the effective mass of electron/hole along the $\beta$ direction ($m^{\beta}_{e(h)}$) can be obtained by calculating the band effective mass at the CBM/VBM
\begin{equation} \label{ee8}
\frac{1}{m^{\beta}_{e(h)}}=\frac{\partial^2\varepsilon}{\hbar^{2}\partial (k^{\beta})^2}|_{\varepsilon=\varepsilon_{CBM(VBM)}},
\end{equation}
where $\epsilon$ and $k^{\beta}$ stand for the energy of band edge and the wave vector along the $\beta$ direction.

We define three types of DP constants
\begin{equation} \label{ee9}
\lambda_{e(h)}^X =\frac{\mathrm{d} \varepsilon }{\mathrm{d} \delta^X } |_{\varepsilon =\varepsilon_{CBM(VBM)}},
\end{equation}
where $\delta^X$ represents different types of strains ($X=L,S\ or\ V$ for the uniaxial, biaxial or volumetric strain), $\varepsilon_{CBM(VBM)}$ denotes the energy at the CBM/VBM
\begin{align}\label{ee10}
\delta^L=\frac{L-L_{0}}{L_{0}}, \\
\delta^S=\frac{S-S_{0}}{S_{0}}, \\
\delta^V=\frac{V-V_{0}}{V_{0}},
\end{align}
where $L_{0}$, $S_{0}$ and $V_{0}$ ($L$, $S$ and $V$) are length, area and volume of crystals without (with) the strains. Note that the DP constant  $\lambda^{\beta}_{e(h)}$ can be obtained by applying the uniaxial strain, and thus is equivalent to the $\lambda^{L}_{e(h)}$. Also, we call $\lambda^X_{e(h)}$ ($L$, $S$ and $V$) as the uniaxial, biaxial and volumetric DP constant, respectively.

Usually, the CBM/VBM of the crystal with high symmetry has many equivalent points that possess the same energy and commonly determine the transport properties. The points can be divided into two categories (copoint or non-copoint degeneracy) based on whether they are at the same position. For instance, the n-type single crystal silicon has six non-copoint degenerate CBMs which symmetrically locate at the \emph{a}*, \emph{b}* and \emph{c}* axes of the reciprocal space. Each two CBMs on the same axis have the same electronic structural properties along the same direction.

Each CBM has an ellipsoidal isoenergetic surface near it, and independently participates in electrical transport. Moreover, each CBM exhibits anisotropic transport behavior because the effective mass varies along different directions of the ellipsoidal isoenergetic surface. However, the total transport of the six CBMs is isotropic. For \emph{i}th CBM, we assume that its contribution to electrical conductivity along one special $\beta$ direction ($\sigma_i^\beta$) is equivalent to the contribution of a spherical isoenergetic surface with the effective mass of the ellipsoidal isoenergetic surface along the $\beta$  direction, rather than the assumption of constant relaxation time. Thus, the total electronic conductivity along the $\beta$ direction can be expressed as
\begin{equation} \label{ee11}
\sigma^\beta=\sum_{i=1}^{N_v}{\sigma^\beta_i}=\sum_{i=1}^{N_v}{n^\beta_{i}qu^\beta_{i}}
={nqu^\beta},
\end{equation}
where $N_V$, $u^\beta$ and \emph{n} are the number of band degeneracies at band edge, the total real mobility and the total real carrier density that is described by
\begin{equation} \label{ee12}
n=\sum_{i=1}^{N_v}{n_{i}}=\frac{V_c}{2\pi^2}(\frac{2m_i}{\hbar^2})^
\frac{3}{2}\varepsilon^{\frac{1}{2}}f,
\end{equation}
Here, $n_i$ is the real carrier density supported by the \emph{i}th band degeneracy, $\varepsilon$, $V_{c}$ and $f$ are the energy of band edge, the volume of unit cell and the Fermi-Dirac
distribution function, and $m_i$ is the local DOS effective mass of \emph{i}th band degeneracy. The $m_i$ is determined by the shape of the isoenergetic surface near the band edge.
The $m_i$ of \emph{i}th spherical isoenergetic surface is band effective mass at CBM/VBM, which is equal to the effective mass along arbitrary $\beta$ direction ($m_i^\beta$). For elliptical isoenergetic surface, it can be written as
\begin{equation} \label{ee13}
m_i=\sqrt[3]{m_i^x m_i^y m_i^z},
\end{equation}
where $m_i^x, m_i^y$ and $m_i^z$ are the effective masses along the three principal axes of the ellipsoidal isoenergetic surface. The $m_i$ is the result of equating the volume of an ellipsoid to a sphere. Mathematically speaking, it is the geometric mean of the effective masses along the three direction. The $m_{i}$ expression for the case of anisotropy has significant limitations. For example, when the effective mass along the one direction is infinitely close to zero, the $m_{i}$ is close to zero. Obviously, this is not in line with the facts. Therefore, it seems more reasonable to use geometric mean of the effective masses to describe the DOS near band edge.
\begin{equation} \label{ee14}
m_i=\frac{m_i^x +m_i^y+ m_i^z}{3},
\end{equation}

In Eqs.(\ref{ee11}), $n^\beta_i$ and $u^\beta_i$ are the equivalent carrier density and the equivalent mobility of \emph{i}th band degeneracy given by
\begin{align}\label{ee15}
n^\beta_i&=\frac{V_c}{2\pi^2}(\frac{2m^\beta_i}{\hbar^2})^
\frac{3}{2}\varepsilon^{\frac{1}{2}}f,
\end{align}

\begin{align}\label{ee16}
\mu_{i}^\beta &=\frac{2\sqrt{2\pi}e\hbar^{4}c^\beta}{3(k_{B}T)^{3/2}(m_i^\beta)^{5/2}(\lambda^{\beta})^2}.
\end{align}
Here, $m^\beta_i$ represents the effective masses of \emph{i}th band CBM/VBM along the same $\beta$ direction. Substituting Eqs. (\ref{ee12}), (\ref{ee15}) and (\ref{ee16}) into (\ref{ee11}), the total real mobility is obtained
\begin{align}\label{ee17}
\mu^\beta &=\frac{2\sqrt{2\pi}e\hbar^{4}c^\beta}{3(k_{B}T)^{3/2}(m_{[m]}^\beta)^{5/2}(\lambda^{\beta})^2}.
\end{align}
We define the $m_{[m]}^\beta$ as the mobility effective mass along the $\beta$ direction,

\begin{align}\label{ee18}
\frac{1}{(m_{[m]}^\beta)^{5/2}}=\frac{\sum\limits_{i=1}^{N_v}\frac{1}{m_i^\beta}}{\sum\limits_{i=1}^
{N_v}(m_i)^{\frac{3}{2}}}.
\end{align}
When the $N_{v}$ band degeneracies have the same isoenergetic surfaces, one can get the more common variation of Eq. (\ref{ee18}),
\begin{align}\label{ee19}
\frac{1}{(m_{[m]}^\beta)^{5/2}}=\frac{\frac{1}{N_v}\sum\limits_{i=1}^{N_v}\frac{1}{m_i^\beta}}{\frac{1}{N_v}\sum\limits_{i=1}^
{N_v}(m_i)^{\frac{3}{2}}}=\frac{\frac{1}{m_c}}{m_i^{3/2}}=\frac{1}{m_cm_i^{3/2}}.
\end{align}
The $m_c$ can be called the conductivity effective mass. In fact, it is the mathematical harmonic mean. For the weak anisotropy, the ratio of $m_c$ and $m_i$ is close to 1. Thus, the other vibration\cite{rr48} of Eq. (\ref{ee18}) is

\begin{align}\label{ee20}
\frac{1}{(m_{[m]}^\beta)^{5/2}}\approx\frac{1}{m_i^{5/2}}.
\end{align}
Moreover, when the same isoenergetic surfaces locate at the same position, intervalley scattering can proceed smoothly like intravalley scattering. Therefore, the Eq.~(\ref{ee16}) should be writen as

\begin{align}\label{ee21}
\mu_{i}^\beta &=\frac{2\sqrt{2\pi}e\hbar^{4}c^\beta}{3N_V(k_{B}T)^{3/2}(m_i^\beta)^{5/2}(\lambda^{\beta})^2}.
\end{align}
and the Eq. (\ref{ee18}) has a variant formula
\begin{align}\label{ee22}
\frac{1}{(m_{[m]}^\beta)^{5/2}}=\frac{\sum\limits_{i=1}^{N_v}\frac{1}{m_i^\beta}}{N_V\sum\limits_{i=1}^
{N_v}(m_i)^{\frac{3}{2}}}=\frac{1}{N_V m_cm_i^{3/2}}
\end{align}

For n-type silicon, when the $\beta$ direction is parallel to the $a^*$ axis, the $m^\beta_1$=$m^\beta_4$, $m^\beta_2$=$m^\beta_5$ and $m^\beta_3$=$m^\beta_6$ correspond to $m_l$, $m_t$ and $m_t$ calling the longitude and transverse effective masses. Six $m_i$ values is equal and it is given by
\begin{align}\label{ee23}
m_i=\sqrt[3]{m_t m_t m_l}.
\end{align}
or
\begin{align}\label{ee24}
m_i=\frac{m_t+ m_t+ m_l}{3}.
\end{align}

Note that the effect of the intervalley scattering on the mobility is not taken account into in n-type silicon, it is because the CBMs possess vastly different momentum (namely locate at different sites), which hinders the intervalley scattering.

Different from the n-type silicon, the p-type silicon has three VBMs converging at $\Gamma$ point. The three degenerate bands contain two heavy bands and one light bands, so $m_i^\beta$ (i=1, 2, 3) is labled by $m_{hh}$, $m_{hh}$ and $m_{lh}$ respectively calling heavy and light band effective masses.
Moreover, the carrier transition occurs between the same momentums and between the same energies, namely, the scattering near the VBMs is approximatively elastic. We assumed that the intervalley scattering is as well as the intravalley scattering. Therefore, the scattering rate is three times the case without intervalley scattering, and the hole mobility of the single crystal is given by
\begin{equation}\label{ee25}
  \mu_{h}^{\beta}=\frac{2\sqrt{2\pi}e
  \hbar^{4}c^\beta}{3(k_{B}T)^{3/2}(m_{[m]h}^\beta)^{5/2}
  (\lambda^{\beta}_{h})^2}.
  \end{equation}
With the help of Eq. (\ref{ee22}), the $m_{[m]h}^\beta$ is roughly expressed as
\begin{align}\label{ee22}
\frac{1}{(m_{[m]h}^\beta)^{5/2}}=\frac{\sum\limits_{i=1}^{3}\frac{1}{m_i^\beta}}{3\sum\limits_{i=1}^
{3}(m_i)^{\frac{3}{2}}}=\frac{1}{3 m_cm_i^{3/2}}.
\end{align}
In fact, its accurate expression is
\begin{align}\label{ee27}
\frac{1}{(m_{[m]h}^\beta)^{5/2}}=\frac{\sum\limits_{i=1}^{3}(m _i^\beta)^{1/2}}{(\sum\limits_{i=1}^
{3}(m_i)^{3/2})^2}.
\end{align}

It is worth mentioning that the $\lambda^\beta$ is replaced by the $\lambda^V$ in the fact calculations because it is difficult to accurately calculate the deformation potentials of different degenerate band edges.
Therefore, the anisotropy of the mobility comes from the contribution of the anisotropy of effective mass and elastic constant that respectively characterize the properties of electrons and phonons.

\section{Results and discussion}
\begin{figure*}[t]
\centering
\includegraphics[width=1.4\columnwidth]{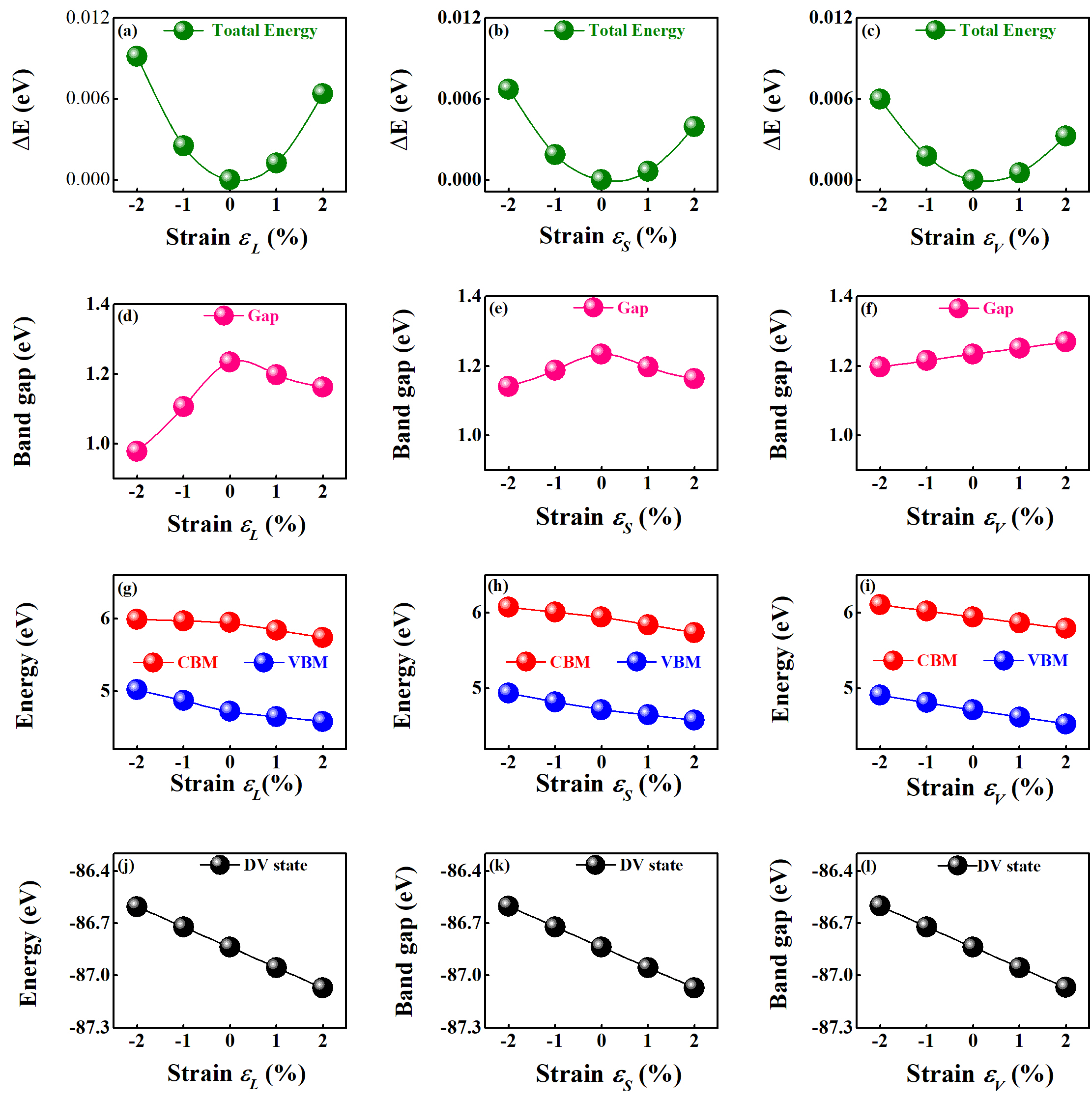}
	\caption{(Color online) The uniaxial/biaxial/volumetric strain ($\varepsilon_L$, $\varepsilon_S$ and $\varepsilon_V$) effects on the total energy (a)-(c), band gap (d)-(f), VBM/CBM energy (g)-(i), and average energy of the DV state (j)-(l).
    \label{Figx1} }
\end{figure*}

The relative total energies $\Delta E$ of the strained silicon with reference to the unstrained structure are all positive values shown in Figs.~\hyperref[Figx1]{\ref*{Figx1}(a)}--\hyperref[Figx1]{\ref*{Figx1}(c)}.
This indicates that the relative error for calculated lattice constant should be less than 1$\%$. It is expected that the calculated lattice constant 5.46 \AA{} for Si agrees well with the experimental value 5.42 \AA{} \cite{rr28}.

The improvements of total energies for the strained structures are related to the breaking of symmetry and the interaction between atoms. Compared to volumetric strain, the uniaxial and biaxial strains have greater impacts on total energy of the system, because the both lead to the reduction of structural symmetry. The uniaxially and  biaxially strained structures belong to space group ($I4_{1}$, No.141), nevertheless, the volumetrically strained structure maintains the same structural symmetry ($Fd\overline{3}m$, No.227) with the unstrained structure.
At the same strain strength, the compression more than the tensility can affect the total energy of silicon, because the repulsive interactions between atoms are more sensitive to the relative positions than the attractive interactions. For instance, the $\Delta E$ at the $-1\%$ uniaxial strain is 0.0025 eV obviously larger than 0.0012 eV at the $1\%$.

The calculated bandgap of unstrained silicon is 1.229 eV, close to the experimental value of 1.17 eV at 0 K \cite{rr29}. The quit small bandgap deviating is mainly due to the calculated lattice constant slightly larger than the experimental data. The good agreement is because the band structure in our work is obtained within the modified Becke-Johnson exchange potential and the generalized gradient approximation for the
correlation (mBJ-GGA) \cite{rr30}, using the full potential method as implemented in the WIEN2k code \cite{rr31}. However, the mBJ-GGA strategy has a shortcoming that no exchange and correlation energy functional is defined \cite{rr41}, so that it cannot be used for all calculations in connection with the total energy of the system, such as total energy and geometric optimization calculations. As a consequence, the total energies in this work are obtained by the standard GGA strategy.

\begin{figure*}[t]
\includegraphics[width=2\columnwidth]{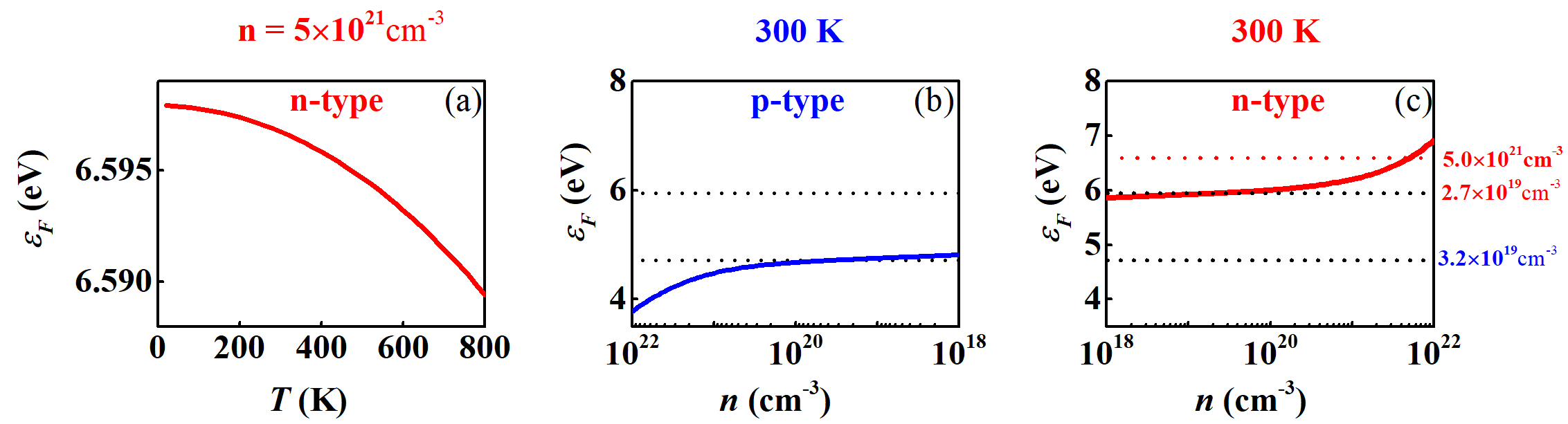}
\caption{(Color online) (a) Temperature dependence of the Fermi level $\varepsilon_F$ at fixed n-type doping density(5$\times$$10^{21}$$cm^{-3}$). (b),(c) Dependence of the Fermi level $\varepsilon_F$ on the p- and n-type doping density at 300 K.
		\label{Figx2} }
\end{figure*}

It is found in Figs.~\hyperref[Figx1]{\ref*{Figx1}(d)} and \hyperref[Figx1]{\ref*{Figx1}(e)} that under uniaxial/biaxial strain the bandgap is positively related to the compressive volume, and negatively to the expansive volume. It is because both the compressive and tensile strains have different effects on the conduction and valence bands. However, the compressive and tensile strains have almost similar effects on the band edge so that the bandgap and volumetric strain are positively correlated throughout the overall region (see Fig.~\hyperref[Figx1]{\ref*{Figx1}(f)}).
Without any energy corrections in Figs.~\hyperref[Figx1]{\ref*{Figx1}(g)}--\hyperref[Figx1]{\ref*{Figx1}(i)}, we found the energy of the band edge goes down as the volume increases, and the standard linear relationship between the band edge energy and the volumetric tensile/compressive strain. However, the CBM energy is more sensitive to the uniaxial tensile strain, on the contrary, the VBM energy is more sensitive to the uniaxial compressive strain, which is also the case in the biaxial strain. The corresponding DP constants are summarized in Table \ref{tx1}. Under compressive strain, the $|\lambda_{VBM}^L|$ is as high as 14.90 eV that is
about seven times of $|\lambda_{CBM}^L|$, and is twice $|\lambda_{VBM}^L|$ under tensile strain. Moreover, the overall $|\lambda_{VBM}^X|$ is larger than the overall $|\lambda_{CBM}^X|$. Especially, the difference between the overall $|\lambda_{VBM}^L|$ and $|\lambda_{CBM}^L|$ reaches as large as 4.59 eV.
\begin{table}
\centering
\caption{\label{tx1}%
Summary of the DP constans (in units of eV) for VBM and CBM. The superscripts $L$, $S$ and $V$ represent obtaining DP constants through uniaxial, biaxial and volumetric  strains, respectively. "Compressed" and "Expanded" represent the DP constant obtained through linear fitting of the data from negative strain, positive strain and the both. "1S" and "2S" represent the corrected DP constant with 1S or 2S core state as the reference energy level, respectively.
}
\resizebox{1.0\columnwidth}{!}{
\begin{tabular}{ccccccccc}
\\
\Xhline{1.2pt}
	  &Compressed & Expanded & This work &1S &2S&Other work \\  
\midrule
$\lambda^{L}_{VBM}$ &-14.90 & -6.91 & -10.92 &1.53 & 0.66 &\multirow{3}{*}{\shortstack{VBM:\\-10.2\footnote{Taken from Ref. \cite{rr8}.}, -7.9\footnote{Taken from Ref. \cite{rr9}}, \\2.05\footnote{Taken from Ref. \cite{rr11}} \\ CBM:\\3.1\footnote{Taken from Ref. \cite{rr24}}, -5.6\footnote{Taken from Ref. \cite{rr9}}}}\\ 
$\lambda^{L}_{CBM}$ &-2.09 &-10.52 & -6.33 & 6.11& 5.25 \\
$\lambda^{S}_{VBM}$ &-10.96 &-6.80 & -8.85 &3.60& 2.73 \\
$\lambda^{S}_{CBM}$ & -6.27  & -10.38 & -8.31&4.14&3.27\\   	
$\lambda^{V}_{VBM}$ &-9.73 &-9.36 & -9.54&2.90&2.05  \\
$\lambda^{V}_{CBM}$ &-7.82 &-7.67 & -7.76&4.68&3.93 \\
\Xhline{1.2pt}
\end{tabular}
}

\end{table}
Previous studies typically used the core state energy as the reference energy level to correct the band edge energy. In pseudopotential method, the use of pseudopotential to handle core electrons makes it impossible to obtain accurate core state energy. Therefore, the deep valence (DV) state energy is adopted as the reference energy level \cite{rr42}, and the DV state energy and the band edge energy are obtained by the analyses of the density of states (DOS) and the band structure along the high-symmetry points containing VBM and CBM, respectively. Due to differences in calculation methods and parameters, the DOS and band energies
usually do not correspond perfectly. This increases the uncertainty of the calculation results on the DP constants. In this work both the DV state energy and the band edge energy are from the self-consistent calculation with very dense \emph{k} points (30000 points in total). We take the average energy of the lowest energy band as the DV state energy. Figs.~\hyperref[Figx1]{\ref*{Figx1}(j)}--\hyperref[Figx1]{\ref*{Figx1}(l)} show the DV state energy and the strain have a negative linear relationship. The influence of the strain types on the DV state energy can be almost ignored owing to small differences between the $\lambda_{dv}^L$, $\lambda_{dv}^S$ and $\lambda_{dv}^V$ (-11.66 eV, -11.68 eV and -11.69 eV). More interestingly, the $\lambda_{dv}^X$ values under the compressive and expanded strains are almost the same, and the difference between them is less than 0.5 eV.

It is worth mentioning that we settle on the full potential rather than pseudopotential method for electronic structure calculations and thus easily obtain the DP constants of core states (1S and 2S) summarized in Table \ref{tx2}. The core state energy like the DV state energy is insensitive to strain types (see Fig.S1 of the Supporting Information), and the differences between them are less than 0.02 eV.
\begin{table}[h]
\centering
\caption{\label{tx2}%
Summary of the calculated uniaxial/biaxial/volumetric DP constans (in units of eV) of core states (1S and 2S).
	}
		\begin{tabular}{cccccc}
\Xhline{1.2pt}
	Core state  &$\lambda_{core}^L$ &$\lambda_{core}^S$  &$\lambda_{core}^V$\\  

\midrule
1S &  -12.4596 &  -12.4592 & -12.4482 \\ 
2S &  -11.5886 & -11.5873  & -11.5954 \\
\Xhline{1.2pt}
\end{tabular}
\end{table}

After correcting the DP constants at the VBM and CBM using the core state energy, we have found some abnormal conclusions as follows: (i) the DP constants change from negative to positive values; (ii) the ratios of $\lambda^X_{VBM}$ to $\lambda^X_{CBM}$ increase significantly; (iii) the $|\lambda^X_{CBM}|$ values are much smaller than those of the $|\lambda^X_{VBM}|$, which is the opposite situation before the correction, because of the reference energy levels. In addition, different strain types with the same strength lead to the difference that is comparable to, even larger than the corrected DP constants. For instance, the difference between the $\lambda^S_{VBM}$ and $\lambda^L_{VBM}$ is as large as 2.07 eV significantly larger than the $\lambda^{L[1S]}_{VBM}$ and $\lambda^{L[2S]}_{VBM}$.
More importantly, choosing different core state energies as the reference levels makes a difference to the DP constants, leading to a large dissimilarity about 1 eV. Evidently, this means that the strain has great influence on the deep core states.

Surprisedly, the corrected DP constants of the CBM, as well as the other theoretical values, are in agreement well with  the experimental data \cite{rr13} that was determined in the heavily As-doped Si with the n-type carrier density as high as 5$\times10^{21}cm^{-3}$.
Unfortunately, the theoretical works neglect an important experimental parameter of the carrier density related with the Fermi energy level. Based on Boltzmann theory, the Fermi energy level corresponds to the carrier density $n$ that is given by
\begin{equation}\label{ee28}
n=n_0-\int{f(\varepsilon,\varepsilon_F,T)D(\varepsilon)d\varepsilon},
\end{equation}
where $\emph{D}$, $\varepsilon_F$ and $\emph{n}_0$ are the total density of states, the Fermi energy level and the number of the valence electrons in the unit cell. Fig.~\hyperref[Figx2]{\ref*{Figx2}} shows that the Fermi
energy level of silicon at the fixed carrier density is insensitive to temperature, however, is highly sensitive to the carrier density at the fixed temperature. When the carrier density changes from $2.7\times10^{19} cm^{-3}$ to $5\times10^{21} cm^{-3}$, the Fermi energy level rises from 5.94 eV (corresponding to CBM) to 6.60 eV. This implies that the experimental DP constants in Ref.~\cite{rr13} is for the Fermi energy level rather than for the band edge where it is theoretically predicted in Refs.~\cite{rr14, rr15, rr16}. Therefore, the high degree of agreement between theoretical and experimental data strongly indicates that the theoretical predictions are problematic. This is also confirmed by the fact that the calculated mobility with the corrected DP constant severely deviates from the experimental data.
\begin{figure}[t]
\includegraphics{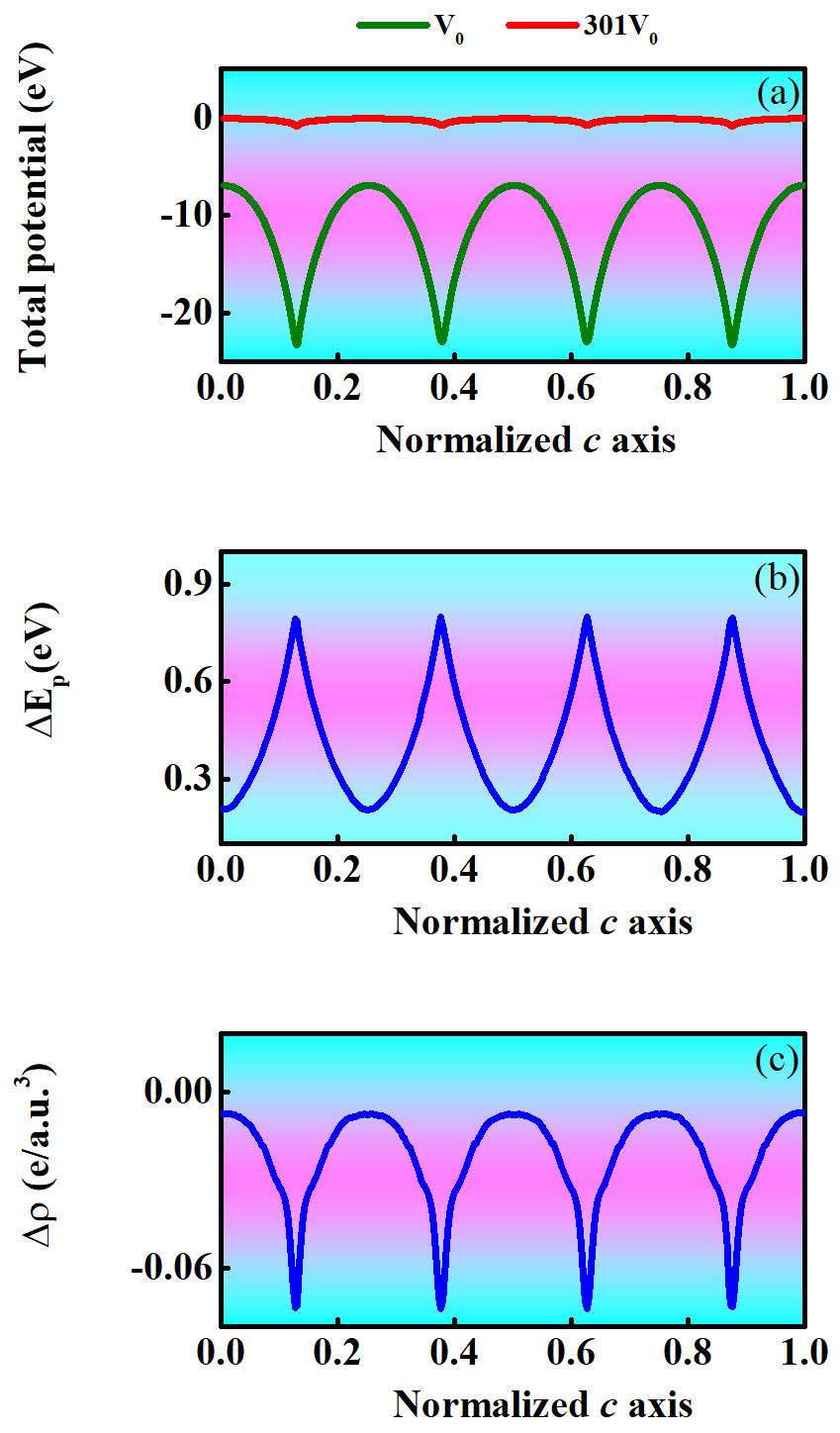}
\caption{(Color online) (a) Plane-averaged potential along the (001) orientation of silicon at V=$V_0$, 301 $V_0$. (b) The difference between the plane-averaged potentials, (c) the difference between the plane-averaged charge density at $V$ = $V_0$ and $V$ = 1.01 $V_0$ under volumetric strain.
\label{Figx3}}
\end{figure}

In this work we consider that the strain affects not only core electrons but also valence electrons because the strain affects the entire periodic potential especially near the silicon nucleus. Fig.~\hyperref[Figx3]{\ref*{Figx3}(a)} shows that  both the strength and shape of the periodic potential have undergone significant changes in the entire region when the volume $V_0$ is expanded to 301 $V_0$. For instance, the periodic potential strength at the $V_0$ is lower than -6.95 eV, however, it rises to roughly 0 eV at the 301 $V_0$, and the potential shape change from the curve like the spring to an approximate straight line. In fact, the two periodic potential values are not absolute due to the different zero-potential points. However, the potential difference can actually reflect the change in the periodic potential. The potential difference in Fig.~\hyperref[Figx3]{\ref*{Figx3}(b)} shows the shape and strength of the periodic potential have changed even if a small volume change of 1\% occurs. Especially, the closer to the silicon nucleus, the greater the periodic potential changes.
\begin{table}[h]
\centering
\caption{\label{tx3}%
	The calculated effective mass of the silicon at different step sizes, the corresponding mean values from averaging the last five step sizes, and the experimental, theoretical data. In addition, the calculated and experimental mobilities at 300 K are listed. The effective mass is in units of the electron rest mass ($m_0$), the mobility is in units of $cm^{2}$/Vs, and the step size is in units of 2$\pi$/4000$a_0$.
	}
 \begin{threeparttable}[b]
\resizebox{1\columnwidth}{!}{
\begin{tabular}{ccccc}
\\
\Xhline{1.2pt}
  Step size& $m_l$ & $m_t$ & $m_{hh}$ &$m_{lh}$ \\
\midrule
10&0.565& 0.141& 0.306& 0.141   \\
20&0.794& 0.204& 0.299& 0.206   \\
40&0.913& 0.209& 0.335&  0.202  \\
50&0.911& 0.211& 0.341&  0.201  \\
80&0.934& 0.214& 0.342& 0.201   \\
100&0.969& 0.214& 0.340& 0.200  \\
200&0.957& 0.216& 0.339&  0.201  \\
\midrule
This work &0.937& 0.213& 0.339&  0.201 \\
Experiment \footnote{The $m_t$ and $m_l$ values are taken from Ref. \cite{rr38}, the $m_{hh}$ and $m_{lh}$ values from \cite{rr32}.} &0.9163& 0.1905& 0.43&  0.19\\
Theory \tnote{2}\footnote{Taken from Ref. \cite{rr39}.} &0.96& 0.16& 0.26&  0.18\\
\Xhline{1.2pt}
 \multirow{2}{*}{Temperature}
 &\multicolumn{2}{m{30mm}}{~~~~~Calculated} &\multicolumn{2}{m{30mm}}{~~~~~~~~~~~~Measured}\\
  &$\mu_p $ & $\mu_n$ & $\mu_p \footnote{Taken from Ref. \cite{rr37}.}$ & $\mu_n \footnote{Taken from Ref. \cite{rr40}.}$\\
  \midrule
300 K &706, 770 & 1771, 2631 & 360 to 510 & 1350 to 1750  \\
\Xhline{1.2pt}
\end{tabular}}
\footnotetext {The $m_t$ and $m_l$ values are taken from Ref. \cite{rr38}, the $m_{hh}$ and\\ $m_{lh}$ values from \cite{rr32}.}
\end{threeparttable}
\label{table1}
\end{table}

The electron density $\rho$(\textbf{r}) is determined by the probability density related to the wave function that is independent of the zero-potential point. The electron density difference truly reflects that the electron density around the Si nucleus has more obvious change (see Fig.~\hyperref[Figx3]{\ref*{Figx3}(c)}). This change originates from the atomic interaction change caused by volume variation, and inevitably leads to the change for the core state energy. Generally, the energy of the core electron is determined by the zero-potential point and the atomic interactions composed of the exchange correction interaction and the Coulomb interaction.

In this work we expand the cell volume large enough so that the effect of the atomic interactions on the core electrons can be neglected. Obviously, the change in the core state energy is mainly due to the zero-potential rather than atomic interactions. Therefore, the correction for DP constant is obtained easily.

\begin{figure}
	\includegraphics[width=1\columnwidth]{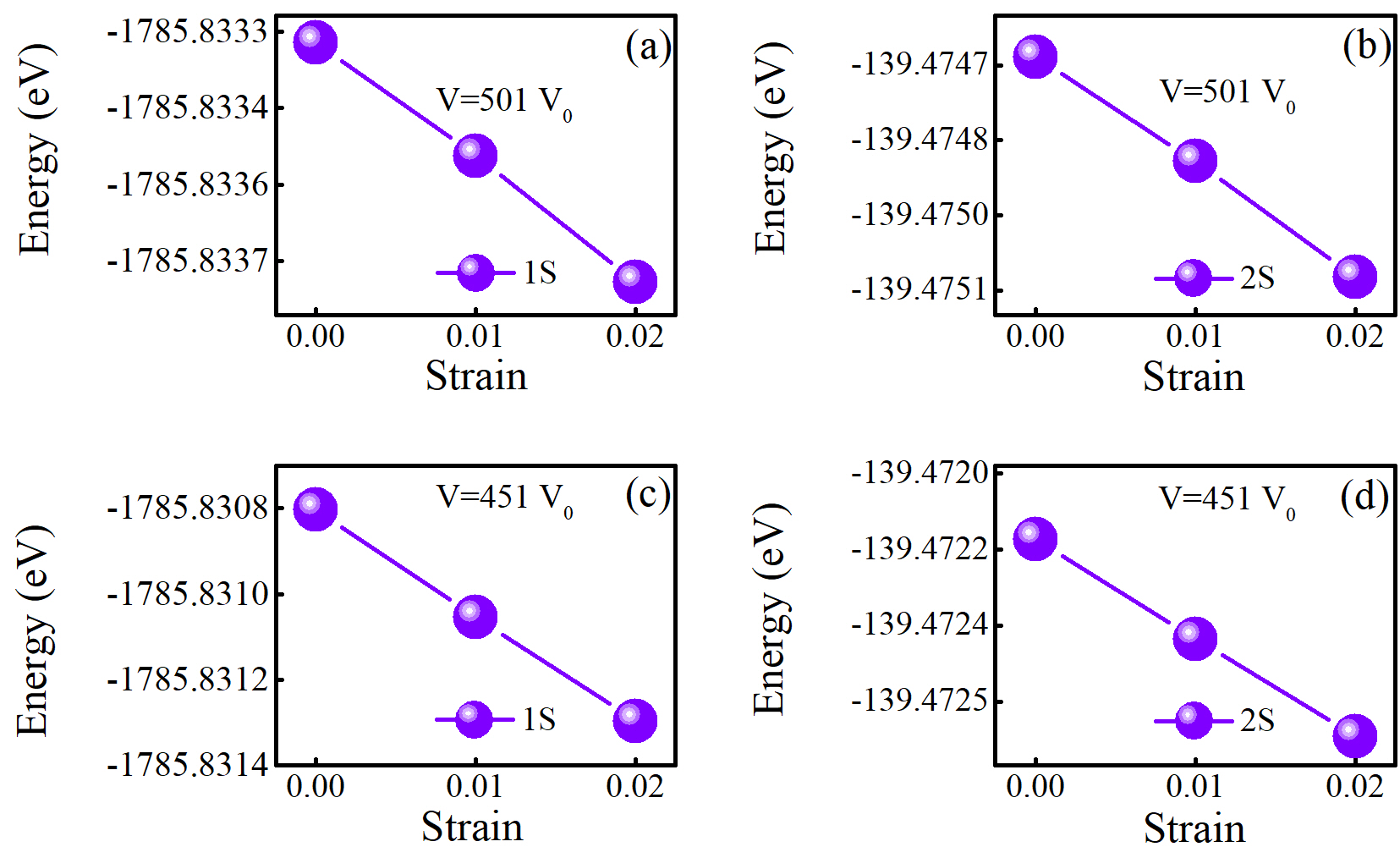}
	\caption{(Color online)  Volumetric strain dependence of 1S and 2S core states (a), (b) at $V$ = 501 $V_0$, and (c), (d) at $V$ = 451 $V_0$.
		\label{Figx4} }
\end{figure}

Considering limited computing resources, we used the standard GGA rather than mBJ-GGA strategy
to perform the self-consistent calculation with rigorous charge and energy convergence limits ($10^{-6}$ e and $10^{-6}$ eV). It is worth noting to ensure the $R_{mt} \cdot K_{max}$ value in all calculations is 7.0 by setting suitable dimension parameters of WIEN2k code. Fig.~\hyperref[Figx4]{\ref*{Figx4}} indicates the relationship between the core state energy and the volumetric strains seems to be linear. The slope values for 1S and 2S states are -0.024601 eV and -0.024605 eV at 451 $V_{0}$, and decrease down to -0.021944 eV and -0.021987 eV at 501 $V_{0}$, respectively.

We consider the slight slope results from the zero-potential change induced by the volumetric strain, and assume that the zero potential $E^{0}$ and the 1S core state energy $E^{1S}$  are approximately equal, and have the property of Coulomb potential
\begin{equation}\label{ee29}
E^{0} \propto \frac{k}{r}.
\end{equation}
Therefore, when the volume change from $V_{N} =N V_{0}$ to (1+$\delta^V$) $V_{N}$, the zero potential difference $\Delta E^{0}_{N}$ can be described by
\begin{equation}\label{ee30}
\Delta E^0_{N} \propto \frac{k}{\sqrt[3]{(1+\delta^V)V_{N}}}-\frac{k}{\sqrt[3]{V_{N}}},
\end{equation}
So the correction for the DP constants at $V_{N}$ is about
\begin{equation}\label{ee31}
\lambda_N^V=\frac{\Delta E^0_{N}}{\delta^V} \approx \frac{\Delta E^{1S}_{N}}{\delta^V}.
\end{equation}
And then one can obtain the correction for the DP constant at $V_{0}$ according the equation
\begin{equation}\label{ee32}
\lambda_1^V=\frac{\Delta E^0_{1}}{\delta^V} \approx \frac{k}{\sqrt[3]{(1+\delta^V)V_{0}}}-\frac{k}{\sqrt[3]{V_{0}}} \approx \sqrt[3]{N} \lambda^V_N
\end{equation}
Herein, taking N=501, the $\lambda_1^V$ is only about -0.1743 eV that is insignificant compared to the $\lambda^V_{e(h)}$ without the correction. It is worth mentioning that we also calculated the slope values at other large volumes, and found that most of them follow this trend of the slope value decreasing with the volume. However, a few slopes do not follow this trend, implying that the uncertainty of zero-potential, but it is certain that the corresponding $|\lambda_1^V|$ is not beyond 0.3 eV.

In addition, the effective mass has an important effect on the mobility, but it is difficult to calculate because it is sensitive to the step size during the numerical derivative process. The dependence of the effective mass with step size for is summarised in Table \ref{tx3}. We average the last five values as the effective mass values. The electron effective mass as well as the light-hole mass is in good agreement well with measured and other calculated values. The heavy-hole mass 0.34 $m_0$ is obviously larger than the other theoretical values, however, it is much closer to the experimental value 0.43 $m_0$ \cite{rr32}.
We adopted the calculated elastic constant $c_{11}$ = 153 GPa that is also in good agreement with experimental and theoretical values \cite{rr33,rr34,rr35,rr36}. Substituting the relevant parameters into the Eqs. (\ref{ee17}) and (\ref{ee25}) and using the local DOS effective masses calculated by Eqs. (\ref{ee14}) and (\ref{ee13}) respectively, the hole mobility at 300 K is calculated to be 704 cm$^{2}$/Vs and 770 cm$^{2}$/Vs that reasonably agrees with experimental value of 510 cm$^{2}$/Vs \cite{rr37} (See Table \ref{table1}). Adopting the mobility effective mass calculated by Eq. (\ref{ee27}), we obtain 1771 $cm^{2}$/Vs that agrees well with experimental data.

\section{Conclusion}
In summary, we improve the mobility formula based on the DP theory, and combine with the first-principle method to calculate the electron and hole mobilities for single crystal silicon. In the improved mobility formula, the mobility effective mass is redefined, more importantly, a method is proposed and applied to calculate the corrections for the DP constant. It is found that the corrections for the DP are slight and thus can be neglected. Our data on the mobility of the single crystal silicon is in good agreement with experimental results, and thus the method can be effectively applied to the prediction of the mobility in bulk materials.\\

\section*{Acknowledgments}
This work is supported by the Scientific and Technological project of Nanchang Institute of Science and Technology (Grant No. NGKJ$-$22$-$09).
\bibliographystyle{aapmrev4-2}
\bibliography{apsguide4-2}

\end{document}